\documentclass[11pt,dvips]{article}

\usepackage{epsfig,times} 
\usepackage{picinpar}
\usepackage{wrapfig}
\usepackage{floatflt}
\makeatletter
\renewcommand{\section}{\@startsection{section}{1}{0in}
	{0.4\baselineskip}{0.1\baselineskip}{\Large\bf}}
\renewcommand{\subsection}{\@startsection{subsection}{2}{0in}
	{0.25\baselineskip}{-\baselineskip}{\large\bf}}
\renewcommand{\subsubsection}{\@startsection{subsubsection}{3}{0in}
	{0.1\baselineskip}{-\baselineskip}{\normalsize\bf}}
\makeatother
\begin{document}

%
\makeatletter\newcommand{\ps@icrc}{
\renewcommand{\@oddhead}{\slshape{HE.6.4.04}\hfil}}
\makeatother\thispagestyle{icrc}
%
%

\begin{center}
%
{\LARGE \bf The Luminescent Bolometer As a Dark Matter Detector} 
\end{center}

\begin{center}
%
%
{\bf L. Gonzalez-Mestres$^{1,2}$}\\
{\it $^{1}$Laboratoire de Physique Corpusculaire, Coll\`ege de France, 75231 Paris Cedex 05, France\\
$^{2}$L.A.P.P., B.P. 110, 74941 Annecy-le-Vieux Cedex, France}
\end{center}

\begin{center}
{\large \bf Abstract\\}
\end{center}
\vspace{-0.5ex}
%
%
Direct detection of WIMP dark matter candidates has to face many
difficult challenges. In particular, it requires an extremely high level of
background rejection. The only way out seems to be particle identification
which, for experiments based on nucleus recoil, is most efficiently performed
by simultaneously detecting ionization or light and phonons. When comparing
different approaches, it is necessary to keep in mind the potential 
requirement of building large detectors and the difficulties that this
condition may raise for some cryogenic devices. It is claimed that the
luminescent bolometer (simultaneous detection of light and phonons) red by
arrays of superconducting tunnel junctions, as proposed by the author some
years ago, ultimately provides the most appropriate WIMP detector. Solar 
neutrino detection and other applications are also briefly discussed.

\vspace{1ex}

%
%
\section{The Luminescent Bolometer}
\label{lumbol.sec}
Simultaneous detection of light and phonons in a single crystal scintillator
cooled to very low temperature
was proposed (Gonzalez-Mestres, \& Perret-Gallix, 1988a) 
as a new tool for high-performance particle detection, expected to:
a) provide slow thermal detectors with a fast light
strobe, giving a much better timing without crucially spoiling energy
resolution; b)
make possible particle identification through the phonon/light ratio,
thus improving background rejection.
As bolometers evolve towards the detection of nonequilibrium
phonons, fluorescence appears as a natural
complement. Many crystals are expected to produce an important light yield at
very low temperature.
The device operating simultaneous detection of light and phonons at very low
temperature was called the $luminescent$
$bolometer$ (Gonzalez-Mestres, \& Perret-Gallix, 1988b).
Simultaneous detection of fluorescence light and nonequilibrium phonons would
combine (Gonzalez-Mestres, 1991a and 1991b) good
energy resolution, fast timing, position information and high background
rejection.
Several extremely difficult experiments may become
feasible, and others 
would be seriously improved.

A number of well-known scintillators (BGO, CdWO$_4$ , CaWO$_4$ , GSO:Ce,
CeF$_3$ , YAG:Ce, CaF$_2$:Eu...) exhibit fast luminescence
at low temperature and are candidates for absorbers.
But other substances, which do not scintillate at room
temperature, become efficient luminophores when cooled down.
Two examples:

a) PbMoO$_4$ green fluorescence is known (Bernhardt, 1985; Van Loo,
\& Wolterink, 1974) 
to increase by four orders of magnitude between room temperature
and LN$_2$ temperature. It has been studied down to
He$_4$ temperature (Van Loo, \& Wolterink, 1974). 
Following a proposal
to use cooled PbMoO$_4$ in double beta experiments (Gonzalez-Mestres,  
\& Perret-Gallix, 1989a), a PbMoO$_4$ $2
\times 2 \times 2$ cm$^3$ single crystal
read by a photomultiplier through a quartz light guide was
characterized down to LN$_2$ temperature and showed a
photopeak pulse height equal to 16$\% $ of that of room temperature
NaI:Tl (M. Minowa et al., 1992).

b) Some indium oxides studied by J.P. Chaminade (see Gaewdang, T., 1993) 
following a proposal (Gonzalez-Mestres, \& Perret-Gallix, 1987 and 1989b)
to develop scintillating single
crystals of In compounds,
exhibit scintillation at low temperature.
In$_2$Si$_2$O$_7$ has been characterized down to 
4K (Gaewdang, 1993), but 
other compounds can be considered (Gonzalez-Mestres, \& Perret-Gallix, 
1987; Gaewdang, 1993).

If the properties of the absorber (Debye temperature, phonon propagation,
low temperature scintillation, light yield and decay
time...) are crucial to the quality of a luminescent bolometer, the sensors
are equally key elements. The first design of the device
proposed (Gonzalez-Mestres, \& Perret-Gallix, 1988 a and 1988b)
the use of separate sensors for the light and the phonons.
Semiconductors,
thin black bolometers and superconductors were considered as photon sensors
(Gonzalez-Mestres, \& Perret-Gallix, 1988a, 1988b and 1989a). 
The first successful feasibility study, made by the Milano group,
adopted such an approach (Alessandrello et al., 1991 and 1992) using
a photodiode as the cryogenic photosensitive device. 
The results were naturally limited in
threshold and energy resolution, as has been discussed in more recent reviews
(Fiorini, 1993; Sadoulet, 1993)
but the situation can be technically
improved. In 1991 , we proposed (Gonzalez-Mestres,
1991a and 1991b) 
a new design incorporating a common sensor for both the light strobe and
the delayed phonon signal, and suggesting also the
use of arrays of superconducting
tunnel junctions (STJ) as the new sensor.

\section{Superconducting Sensors}
\label{supersen.sec}

Superconductors are natural sensors for both phonons and photons, and should
perform better than semiconductors due to
the comparatively low gap for quasiparticle excitation.
Low impedance superconducting films already provide the best phonon sensors
(f.i. Ferger, 1994),
and photosensitive superconducting devices are an active research
subject (f.i. Barone, \& Russo, 1993).
A performant superconducting sensor, sensitive to the light strobe
followed by the delayed pulse of phonons,
would considerably simplify and improve the architecture of the luminescent
bolometer.
This seems feasible nowadays due to the success of arrays of superconducting
tunnel junctions (f.i. Goldie, 1990) and of other
superconducting sensors. The new device,
made of a single crystal low temperature scintillator with an appropriate
superconducting sensor implanted on each of its faces,
may become the {\it ultimate} detector for several physics goals 
(Gonzalez-Mestres, 1991b).

\subsection{STJ Arrays}
\label{stj.sec}
In a very important work, 
with a series array of 432 Al-Al STJ, implanted on a Si wafer with an area of
12 $\times $ 12 mm$^2$ and a thickness of 0.5 mm,
the Oxford group obtained (Goldie, 1990)
at T $\simeq $ 360 mK (base temperature of
the cryostat) a resolution
of 700 eV FWHM on a 25 keV X-ray peak produced from the fluorescence of an
indium foil.
A naive extrapolation suggests (Gonzalez-Mestres, 1991a and 1991b)
that a similar
STJ array could be sensitive to $\approx $ 1 keV of scintillation
photons absorbed near
the array, which corresponds to the light
yield of a $\approx $ 10 keV electron or photon in an efficient scintillator.
According to the Oxford group, most of the 25 keV peak width was due
to drifts in the cryostat temperature
and to electronic noise.
These considerations motivated our 
proposal to use arrays of STJ for the
detection
of both photons (the fast strobe) and phonons (the delayed pulse).
To absorb the scintillation light we considered
(see also Gonzalez-Mestres, 1994), 
either implanting a thin layer between the radiation absorber and the STJ
array (but care must be taken of phonon propagation through the
layer), or to use blackened STJ arrays covering a large fraction of the crystal surface
(which requires working in an optical cavity).

An interesting possibility would be to deposit the STJ arrays on a
superconducting substrate
of higher critical temperature, covering the full crystal surface
(an array per face). Nb or Sn can be the substrate for a Al STJ array.
Thus, both the photons and the phonons from the absorber would be
converted into quasiparticles
by the substrate layer in an efficient way.
Such quasiparticles would subsequently be detected by the STJ array.
The photon signal would immediately originate in the substrate,
whereas phonons would first undergo a number of scattering processes
depending on the size and quality of the crystal.
With an ADC and a DSP after the electronic chain, for each face of
the crystal, digital analysis
would allow to reach a very low threshold for both the fluorescence
and the phonon signal. 
When the expected signal is a sum of exponentials, on-line digital
filtering allows for iterative algorithms
leading to very performant trigger schemes (Gonzalez-Mestres, 1992a),
which apply to fluorescence in a straightforward way and can be adapted
to phonon
detection. Energy resolution would also be very good,
as total energy can be reconstructed from light and phonon pulses.
Digital analysis
of the phonon pulse would lead to excellent space resolution inside the crystal, and we
can expect to reach fast timing (down to 100 ns) with suitable choices.
However, the crystal size and phonon scattering properties will necessarily set an intrinsic limit to the detector performance.
Phonons reaching the crystal surface with an energy E $<$ 2$\Delta $
(the gap of the STJ
superconducting material) cannot contribute to the signal.

Perryman et al. have proposed (Perryman, Foden,  
\& Peacock, 1993) 
optical photon counting with a superconducting
substrate in combination with an array of
widely spaced STJ of lower energy gap. STJ are indeed being successful
in the field (Perryman et al., 1999).
In our case, we are not interested in optical photon counting but in the
detection of 10$^2$ - 10$^5$ optical photons
produced by a particle interacting with the cooled scintillator.
On the other hand,
we must face the extra requirement of efficient phonon detection from large
absorbers.
For the phonon yield, the 
fact that several intrinsic scintillators work at low temperature
is encouraging,
as doped scintillators may exhibit poor phonon propagation.

\subsection{Alternatives}
\label{other.sec}
Superconducting films are very successful and solutions based on
this technique, others than STJ arrays (e.g. transition edge sensors,
see Cabrera et al., 1998),
deserve serious consideration. But efficient detection of scintillation
light is likely to
limit the freedom of the design.
It is possible to consider superheated superconducting dots when
only four faces of the crystal need to be used
(allowing for a magnetic field parallel to the four faces),
or for cylinder structures. 
Sensitivity may, however, be a problem for this kind of detectors.
It seems difficult to simultaneously detect light and external phonons using
superheated microspheres,
but we may hope for technical progress in the interface between the granules
and the scintillating absorber.
As compared to other techniques,
STJ arrays present, for large detectors, the advantage of operating with
excellent
performances at He$_3$ and even at He$_4$ temperatures, which 
considerably simplifies cryogenics with respect to large dilution 
refrigerators which would be difficult to handle. 
Simultaneous detection of light and phonons presents a similar advantage as
compared to simultaneous detection of ionization
and heat.

\section{Proposed Applications}
\label{app.sec}
The luminescent bolometer is an exceptionally versatile device, with many 
possible applications involving large (especially 
if it can be operated at He$_3$
temperature) and small detectors.

\subsection{Non-Baryonic Dark Matter}
\label{dm.sec}
This was the first proposed application of the luminescent bolometer 
(Gonzalez-Mestres, \& Perret-Gallix, 1988a and 1998b), in
view of background rejection and nucleus recoil
identification. The approach was criticized on the grounds of the high 
threshold of existing prototypes,
but as explained above the situation can be
considerably improved introducing superconducting
sensors: then, the threshold of the luminescent bolometer will become as low as
that of any device performing
simultaneous detection of ionization and heat. The possibility to work well
above dilution temperatures will then become
a definite advantage of dark matter experiments using the luminescent
bolometer.
The reliability of a large scale experiment would be much better with our
approach, where 100 kg to 1 ton detectors can indeed
be cooled to the operating temperature with existing and well established
cryogenic techniques. Detector stability would also be a crucial
advantage. Furthermore,
targets such as $^7$Li, $^{19}$F, $^{27}$Al, $^{127}$I, $^{183}$W... can be
incorporated in the cold scintillator approach.
If particle physics and cosmology still provide a ground to experiments aiming
at the direct detection of dark matter WIMPs
(there is to date no evidence for new particles!),
the luminescent bolometer with a superconducting read-out is to be the right
technique for that purpose.
However, many basic studies remain to be performed on the low temperature
behaviour of the relevant scintillators.

\subsection{Double Beta}
\label{db.ps}
Applications to double beta experiments were proposed in 1989 
(Gonzalez-Mestres, \& Perret-Gallix, 1988a), with the
basic idea of rejecting the alpha background in high
Q materials. CdWO$_4$ and PbMoO$_4$ were then explicitly considered.
The use of CdWO$_4$ seems indeed to be a promising way 
(Zdesenko et al., 1992; Alessandrello et al., 1991a), and Mo would be
a very performant target. To our
original proposal, the Milano group has added
the successful development of a CaF$_2$ luminescent bolometer 
(Alessandrello et al., 1991a and 1991b).
After suitable technical developements, the luminescent bolometer can
potentially incorporate any double beta target.
The elementary cells
of a double $\beta $ experiment
can be $\approx $ 2 $\times $ 2 $\times $ 2 cm$^3$ crystals, which amounts to
$\approx $ 25 crystals and
$\approx $ 150 electronic channels per Kg of detector.

\subsection{Solar and Reactor Neutrinos (Indium Target)}
\label{in.sec}
This may become the most important and far-reaching application of the
luminescent bolometer, as no technique allows
by now to detect in real time low energy neutrinos.
Several indium compounds
seem to scintillate mainly at low temperature.
To the indium germanates, silicates and other oxides presently under study,
some of which (e.g. In$_2$Si$_2$O$_7$) give excellent results at
low temperature, InCe oxides should be added in order to possibly
exploit the fluorescence of trivalent cerium. Fluorides deserve further
consideration (Gaewdang, 1993; Gonzalez-Mestres and Perret-Gallix, 1987), 
as some of them can scintillate
and crystal growth seems easier than with oxides.
A large scale, real time solar neutrino experiment
based on Raghavan'as reaction, with a luminescent bolometer made of an indium
compound, has already been described (Gonzalez-Mestres, 1991a, 1991b, 1992b
and 1994)
and nothing to date
contradicts its potential feasibility.
A neutrino-antineutrino oscillation experiment at a reactor would be $\sim $
100 times smaller.
It must therefore be considered as a suitable intermediate step. 

\subsection{Other Applications}
\label{more.sec}
Basic physics and chemistry, (f.i. study of relaxation
phenomena), nuclear physics and technology 
(f.i. nuclear spectroscopy, heavy-ion physics), astrophysics...
are generating potential applications
of the luminescent bolometer (Gonzalez-Mestres, 1994). Neutron 
detection at low counting rate (e.g. with a
lithium target, Gonzalez-Mestres 1991a; de Marcillac et al., 1993), 
low radioactivity measurements, 
long lived isotopes... present 
increasing interest in both scientific and industrial domains. 

%
%
%
\vspace{1ex}
\begin{center}
{\Large\bf References}
\end{center}
%
Alessandrello, A. et al., 1991a, Proceedings of LTD-4,
Oxford September 1991 , Ed. Fronti\`eres, p. 367.\\
Alessandrello, A. et al., 1991b, Proceedings of the Moriond Workshop "Progress
in Atomic Physics, Neutrinos and Gravitation" January
1992 , Ed. Fronti\`eres, p. 201.\\ 
Barone, A., \& Russo, M., 1993, in Advances in Superconductivity, Plenum.\\
Bernhardt, Hj., 1985, Phys. Stat. Sol. (a) 91, p. 643.\\
Cabrera, B., et al., 1998, Appl. Phys. Let. 73(6), 735.\\
Ferger, P., et al., 1994, "A Massive Cryogenic Particle Detector with Good 
Energy Resolution", Max-Planck-Institut
preprint Munich.\\
Fiorini, E., 1993, Proceedings of LTD-5 , Journal 
of
Low Temperature Physics 93, Numbers 3/4 , p.189.\\
Gaewdang, T., 1993, Thesis Universit\'e de Bordeaux
I "Cristallochimie et luminescence de quelques oxides et
fluorures d'indium", November 1993.\\
Goldie, D.J., in Proceedings of the Workshop on Tunnel Junction
Detectors for X-rays, Naples December 1990 ,
World Scientific.\\
Gonzalez-Mestres, L., and Perret-Gallix, D., 1987, Proceedings of the 
"Rencontre sur la Masse Cach\'ee", Annecy July 1987 ,
Ed. Annales de Physique, p. 181.\\
Gonzalez-Mestres, L., and Perret-Gallix, D., 1988a, in Low Temperature 
Detectors for
Neutrinos and Dark Matter - II", Proceedings
of LTD-2 Annecy May 1988 , Editions Fronti\`eres.\\
Gonzalez-Mestres, L., and Perret-Gallix, D., 1988b, Proceedings of the XXIV
International Conference on High Energy Physics,
Munich August 1988 , Ed. Springer-Verlag, p. 1223.\\
Gonzalez-Mestres, L., and Perret-Gallix, D., 1989a, Moriond Workshop "The Quest for
Fundamental Constants in Cosmology", March 1989 ,
Ed. Fronti\`eres, p. 352.\\
Gonzalez-Mestres, L., and Perret-Gallix, D., 1989b, Nucl. Instr. and Meth.
A279, p. 382.\\
Gonzalez-Mestres, L., 1991a, Proceedings of LTD-4,
Oxford September 1991 , Ed. Fronti\`eres, p. 471.\\
Gonzalez-Mestres, L., 1991b, Proc. TAUP 91, Nuclear Physics B
(Proc. Suppl.) 28A (1992), p. 478-481.\\
Gonzalez-Mestres, L., 1992a, unpublished: see S. Dil, Rapport de stage
DESS Universit\'e Jean Monnet, Saint-Etienne June 1992.\\
Gonzalez-Mestres, L., 1992b, Proceedings of the Moriond Workshop "Progress in
Atomic Physics, Neutrinos and Gravitation"
January 1992 , Editions Fronti\`eres, p. 113.\\
Gonzalez-Mestres, L., 1994, paper physics/9711025 of LANL (Los Alamos)
electronic archive.\\
Marcillac, P. de, et al., 1993, Nuclear Instruments and Methods A
337, p. 95.\\
Minowa, M. et al., 1992, University of Tokyo
preprint UT-HE-92/06.\\
Perryman, M.A.C., Foden, C.L., \& Peacock, A., 1993, 
Nuc. Inst. Meth. A 325, 319.\\
Perryman, M.A.C. et al., 1999, paper astro-ph/9905018 of LANL archive.\\
Sadoulet, B., 1993, Proceedings of LTD-5 , 
Journal of
Low Temp. Phys. 93, Numbers 3/4 , p. 821.\\
Van Loo, W. \& Wolterink, D.J., 1974, Phys. Lett. 47A, p. 83.\\
Zdesenko, Y. et al., 1992, Proceedings of the January Moriond Workshop, 
Editions Fronti\`eres, p. 183.
\end{document}